\documentclass[aps,prb,reprint,groupedaddress]{revtex4-1}
\usepackage {graphicx}  % needed for figures (not in this templat)

\usepackage{epstopdf}
\usepackage{bm}        % for math (not in this templat)
\usepackage{amssymb}   % for math (not in this templat)

\makeatletter

\newcommand{\Rmnum}[1]{\expandafter\@slowromancap\romannumeral #1@}
\makeatother

\begin{document}

\title{Disorder broadening of even denominator fractional quantum Hall states \\
in the presence of a short-range alloy potential}

\author{E. Kleinbaum$^1$, Hongxi Li$^1$, N. Deng$^1$, 
G.C. Gardner$^{1,2}$, M.J. Manfra$^{1,2,3}$, and G.A. Cs\'{a}thy$^{1,2}$}

\affiliation{${}^1$ Department of Physics, Purdue University, West Lafayette, IN 47907, USA \\
${}^2$ Birck Nanotechnology Center, Purdue University, West Lafayette, IN 47907, USA  \\
${}^3$ School of Materials Engineering and School of Electrical and Computer Engineering,
Purdue University, West Lafayette, IN 47907, USA \\}

\date{\today}

\begin{abstract}

We study energy gaps of the $\nu=7/2$ and $\nu=5/2$ fractional quantum Hall states
in a series of two-dimensional electron gases containing alloy disorder.
We found that gaps at these two filling factors have the same suppression rate with alloy disorder.
The dimensionless intrinsic gaps  in our alloy samples obtained from the model proposed 
by Morf and d'Ambrumenil are consistent with numerical results, 
but are larger than those obtained from experiments on pristine samples
published in the literature.
The disorder broadening parameter has large uncertainties. However,
a modified analysis relying on shared intrinsic gaps yields consistent results for 
both the $\nu=5/2$ and $7/2$
fractional quantum Hall states and establishes a linear relationship between the disorder
broadening parameter and alloy concentration.
Furthermore, we find that we can separate contributions to the disorder broadening of the 
long-range and short-range scattering. 

\end{abstract}

\maketitle

Disorder is one of the least understood factors impacting many-body
ground states, such as the ones forming in the two-dimensional electron gas (2DEG).
As disorder levels in 2DEGs in GaAs/AlGaAs heterostructures are reduced, 
there is an increasing number of FQHSs developing \cite{tsui,panF,xiaF}. 
The influence of disorder on other many-electron states,
such as different types of electron solids, is more intricate \cite{luhman,moon,zud}.

The importance of quantifying disorder effects on the energy gap of FQHSs
was recognized early on and played a role in establishing
the Laughlin nature of the $\nu=1/3$ FQHS \cite{mod1,mod2,mod3,mod4}. These efforts yielded a 
simple phenomenological model that relates $\Delta^{meas}$, the measured energy gap, to
$\Delta^{int}$, the gap obtained from numerical simulations 
in disorder-free models \cite{mod1,mod2,mod3,mod4}. 
$\Delta^{int}$ is referred to as the intrinsic gap.
According to this phenomenological model, the measured energy gap $\Delta^{meas}$ is reduced as compared to
the intrinsic gap $\Delta^{int}$ by the so called disorder broadening parameter $\Gamma$:
\begin{equation}
\Delta^{meas}=\Delta^{int}-\Gamma.
\end{equation}
More recent theoretical efforts sought understanding of the
disorder-driven collapse of FQHSs\cite{sheng,wan} and of the impact of disorder
on quantum entanglement \cite{bhatt1,bhatt2,barry}. 

Even denominator FQHSs, such as the ones developing at $\nu=5/2$ and $\nu=7/2$
in 2DEGs in the GaAs/AlGaAs system \cite{willett,eis,pan99,eisen},
are also affected by disorder \cite{pan08,choi,dean,umansky,nuebler,nodar,pan11,gamez,deng,reichl,qian}. 
These FQHSs continue to attract 
interest because of their possible non-Abelian excitations \cite{moore,nayak,stern}.
Within the framework of the phenomenological model for the energy gaps presented above, 
in order to extract the intrinsic gap from measurements
one must independently obtain both $\Delta^{meas}$ and
$\Gamma$ from the measured data. This is clearly not possible from a single measurement, 
say from the knowledge of $\Delta^{meas}_{5/2}$, the measured energy gap of the $\nu=5/2$ FQHS. 
Instead, Morf and d'Ambrumenil \cite{ma} proposed an analysis of gaps
based on the measurement of two independent quantities,  
the gaps measured at particle-hole conjugated filling factors $\nu=5/2$ and $\nu=7/2$.
Such an analysis was first applied to 2DEGs
of the highest available mobility \cite{ma}, henceforth referred to
as pristine 2DEGs. The extracted intrinsic gap values
were in reasonable agreement with numerical results 
both in high density samples \cite{ma,nodar}, with electron density
close to $3 \times 10^{11}$~cm$^{-2}$, as well as in samples
of reduced density \cite{nodar}, as low as $8.3 \times 10^{10}$~cm$^{-2}$.

The analysis of Morf and d'Ambrumenil \cite{ma} of the even
denominator FQHSs remains important since efforts to relate energy gaps and disorder
broadening parameters of these states
to lifetime parameters failed in pristine samples. Indeed, it was found that the measured energy gap
of the $\nu=5/2$ FQHS does not correlate in an obvious way with either the transport lifetime 
\cite{pan08,choi,dean,umansky,nuebler,nodar,pan11,gamez,deng}
or the quantum lifetime  \cite{nodar,nuebler,qian}.  In fact in gated samples it was found that the quantum
lifetime is approximately constant over the density range at which the energy gap
at $\nu=5/2$ decreased from its largest value to zero \cite{nuebler,qian}.
The lack of correlation between the energy gap and lifetime parameters
is not surprising since, in contrast to the energy gap at $\nu= 5/2$, 
both the transport and the quantum lifetimes are 
measured at or near zero magnetic field, in a
regime in which single-electron descriptions work well.

Besides pristine samples, the energy gap of the $\nu=5/2$
FQHS was also studied in a series of samples 
with short range scattering centers deliberately introduced
during the Molecular Beam Epitaxy (MBE) growth process \cite{deng}. Since
these short range scattering centers were Al atoms added to the
GaAs channel \cite{wanli0,gardner}, we will refer to these samples as alloy samples.
Since energy gap measurements at $\nu=7/2$ in alloy samples were not
available, the analysis of Morf and d'Ambrumenil for the
even denominator FQHSs in alloy samples so far could not be performed.

In this Article we report energy gap measurements for
the $\nu=7/2$ FQHS in GaAs/AlGaAs based samples with alloy disorder.
We found that gaps at $\nu=7/2$ and  $\nu=5/2$ have the same suppression rate with alloy disorder.
These samples with alloy disorder
provided an opportunity to examine the model put forth by Morf and d'Ambrumenil \cite{ma}
in the presence of a short-rage scattering potential. We found that this model yields intrinsic energy gaps
consistent with numerical results, albeit 
larger than those obtained from pristine samples
published in the literature.
The disorder broadening parameter extracted using this model 
had a significant scatter that resulted in an
unreasonable dependence on the alloy content of samples. 
However, the additional information of shared dimensionless energy gap
can be exploited to determine $\Gamma$ and its alloy content dependence.
This analysis allowed us to separate contributions to
the disorder broadening due to alloy scattering and other scattering mechanisms.
Our results highlight the contrast between effects of short-range and long-range
scattering potentials on the even denominator FQHSs,
further understanding of the energy gaps of these states, 
and open the door for future studies of other types of
limiting disorder, such as charged impurity disorder.

Samples used in this study are the same as those in Ref.\cite{deng}. They are
$30$~nm quantum well samples of nearly the same electron
density, but in which the GaAs channel has Al atoms added
during the MBE growth process in order to form an Al$_x$Ga$_{1-x}$As alloy \cite{gardner}.
Here $x$ is the Al molar fraction which is significantly less than the Al molar fraction in the confining
layers and it therefore does not affect the electronic wave function perpendicular to the 
plane of the 2DEG. Detailed sample structures and characterization can be found in Ref.\cite{gardner}.
Electrical transport measurements were performed 
at dilution refrigerator temperatures in a van der Pauw geometry,
after samples were illuminated
with a red light-emitting diode \cite{deng}. Samples were mounted in a
He-3 immersion cell and the temperature $T$ of the He-3 liquid  was monitored
by quartz tuning fork viscometry \cite{setup}.
%One essential feature of this series of samples is the small variance of electron density of less than 7$\%$.

Fig.1 shows the longitudinal magnetoresistance
$R_{xx}$ and Hall resistance $R_{xy}$ in the range of filling factors $3<\nu<4$
known as the upper spin branch of the second Landau level.
Data are shown for three samples: the pristine sample $x=0$ 
and two alloy samples with $x=0.00057$ and $x=0.00075$. 
Due to its high mobility of $20\times10^6$~cm$^2$/Vs, 
the pristine sample exhibits all known ground states. The most prominent FQHS
is the one at $\nu=7/2$, but there are also others at
$\nu=3+1/5$, $\nu=3+4/5$ \cite{eisen} and $\nu=3+1/3$ \cite{ethan}.
The high quality of the sample also allows us to observe eight
reentrant integer quantum Hall states, ground states
associated with electronic bubble phases \cite{eisen,den}.

\begin{figure}[t]
 \includegraphics[width=.9\columnwidth]{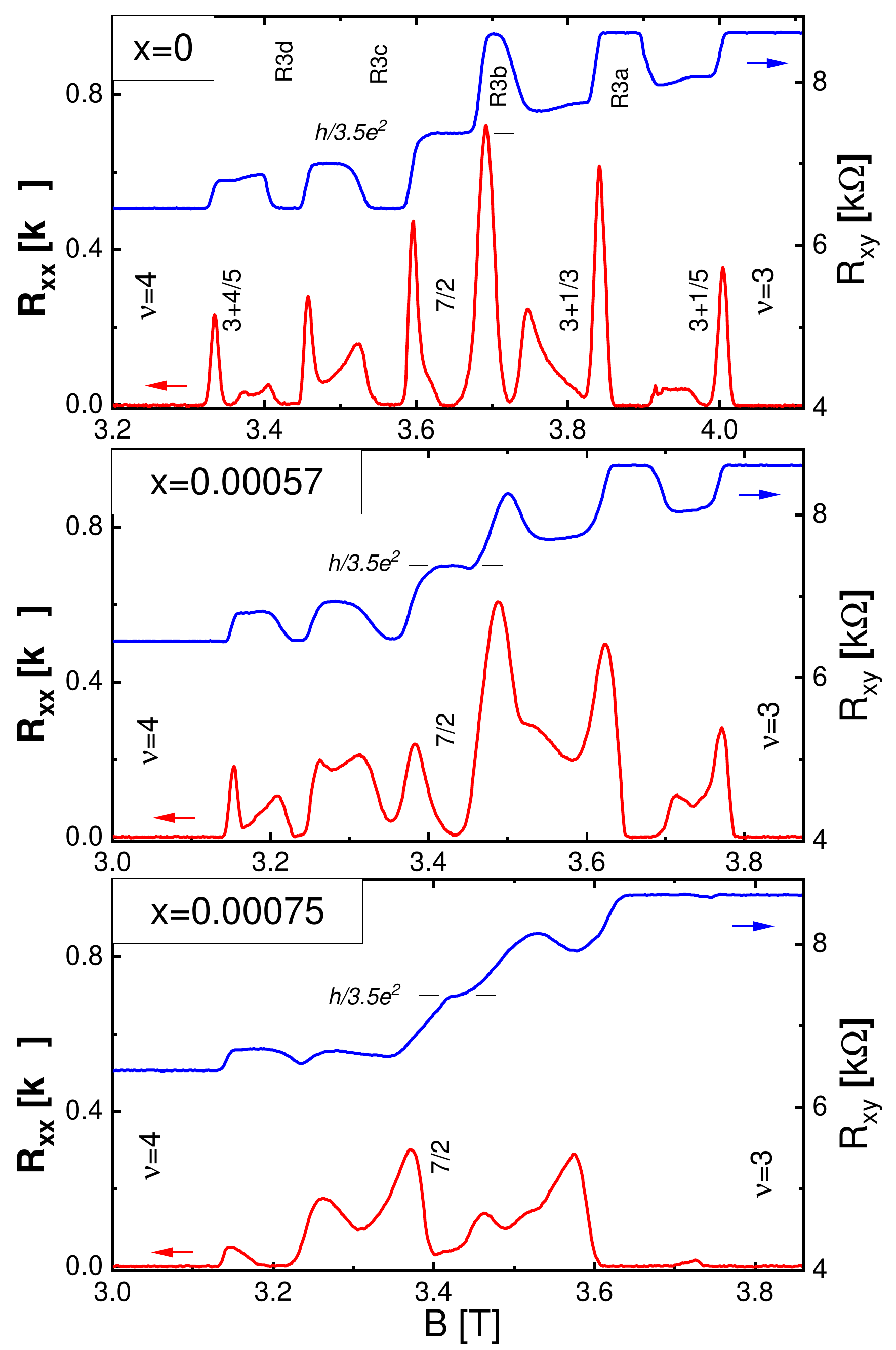}
 \caption{ Magnetoresistance $R_{xx}$ and Hall resistance $R_{xy}$ in 
 samples with $x=0$, $x=0.00057$, $x=0.00075$ alloy content in the filling factor range of $3<\nu<4$ measured at $7$~mK.
 Numbers indicate the filling factors of various FQHSs;  reentrant
 integer quantum Hall states associated with electronic bubble
 phases are labeled $R3a$, $R3b$, $R3c$ and $R3d$.
 \label{Fig1}}
 \end{figure}

An increasing alloy content $x$ has a strong impact on magnetotransport. 
The FQHSs at  $\nu=3+1/5$ and $\nu=3+4/5$
significantly weaken and the one at $\nu=3+1/3$ is destroyed
at the lowest non-zero alloy concentration $x=0.00057$.
Similarly, magnetotransport in the bubble phases is impacted.
However, as observed in both  $R_{xx}$ and $R_{xy}$ data,
the FQHS at $\nu=7/2$ survives
at both $x=0.00057$ and $x=0.00075$, but it is
destroyed in the sample with $x=0.0015$ (not shown). 

In order to characterize the $\nu=7/2$ FQHS, we measure its energy gap. In Fig.2 
we show the temperature dependence of the longitudinal magnetoresistance measured 
at $\nu=7/2$ in the three samples in which a FQHS at this filling factor is present.
As seen in the Arrhenius plots of Fig.2, $R_{xx}$ of all three samples follows an 
activated form $R_{xx} \propto \exp(-\Delta^{meas}_{7/2}/ 2 k_B T)$ at the lowest temperatures. 
Energy gaps $\Delta^{meas}_{7/2}$ extracted 
are $139$~mK, $87$~mK and $23$~mK at $x=0$, $x=0.00057$ and $x=0.00075$, respectively.

\begin{figure}[t]
 \includegraphics[width=0.85\columnwidth]{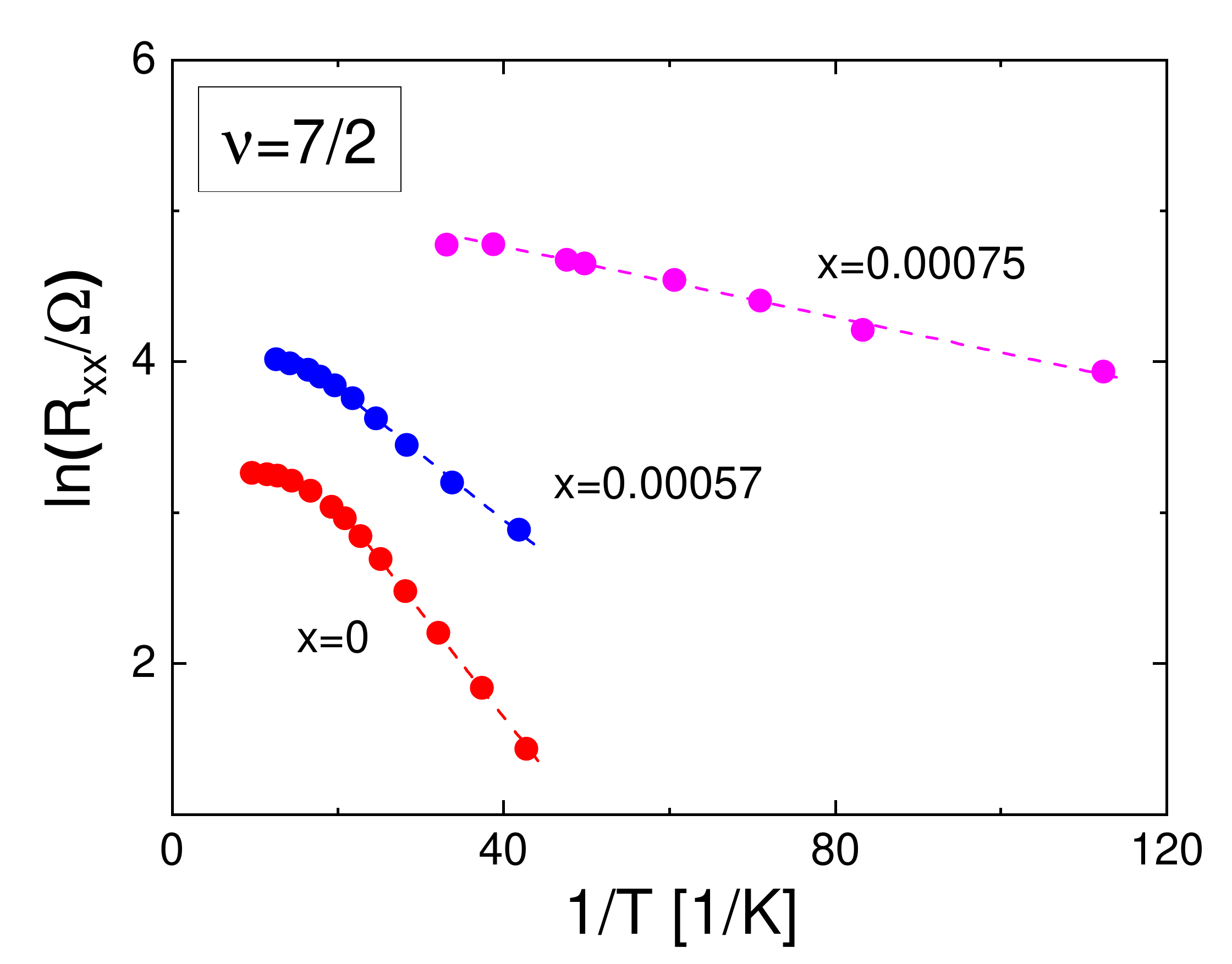} 
 \caption{ Arrhenius plots for the $\nu=7/2$ FQHS at different Al molar fraction $x$. 
 Dotted lines are linear fits in the activated regime used to extract energy gaps.
 \label{Fig2}}
 \end{figure}
 
Similarly to observations at $\nu=5/2$ \cite{deng}, the measured energy gap at $\nu=7/2$
is suppressed by an increasing $x$. Consistent with prior knowledge 
from pristine samples \cite{eisen,dean,nodar}, the energy gap at $\nu=7/2$ in alloy samples is 
also significantly less than that measured at $\nu=5/2$. 
As seen in Fig.3, energy gaps for both FQHSs exhibit a linear trend with the alloy
content $x$; gap suppression rate at both $\nu=5/2$ and $\nu=7/2$ 
are similar, with a value of $\delta \Delta^{meas}/ \delta x \approx 0.15 \times 10^{3}$~K.
This behavior indicates that both  $\nu=5/2$ and $\nu=7/2$ FQHSs respond in a similar
fasion to alloy disorder.

While the two energy gaps plotted against $x$ in Fig.3 exhibit linear trends,
data display significant scatter. Scatter in this figure has two distinct sources:
scatter in the energy gap and scatter in $x$, the Al content of the alloy channel.
Scatter in the energy gap is well-known from measurements in pristine samples. 
For example, after repeated cycling of the sample to room temperature,
the measured energy gap is known to have small varition variations. Similarly, variations in the sample state
due to the sample illumination procedure are likely present. 
Systematic errors in the temperature measurement may also contribute to errors 
in the energy gap. These sources of scatter are often estimated to contribute to 
about $\pm 10\%$ error to the gap.
In addition, some scatter of data in Fig.3 can be associated the alloy forming process.
Because of the extreemely low amount of Al in the channel of our alloy samples,
the Al effusion cell used for the channel alloy formation is operated in a regime of very low flux \cite{gardner}.
Under these conditions there will be errors in $x$, the Al content of the channel.
While such errors are diffcult to quantify, their presence can be observed
as deviations from a linear dependence 
of the scattering rate versus $x$ in Ref.\cite{gardner}. 
We attribute to errors in $x$ the correlated scatter of the energy gaps at 
$\nu=5/2$ and $\nu=7/2$ present at the three lowest $x$ values in Fig.3.

In the following we analyze the energy gaps of the two even denominator FQHSs 
using the model proposed by Morf and d'Ambrumenil \cite{ma}.
Since FQHSs are many-body ground states, the intrinsic energy gap scales with the Coulomb energy $\Delta^{int}=\delta^{int}E_{C}$.
Here $E_C$ is the Coulomb energy $E_C=e^2/4 \pi \epsilon l_B$, $l_B=\sqrt{\hbar/eB}$ is the 
magnetic length. The dimensionless intrinsic gap $\delta^{int}$ depends on the
type of electronic correlations, the thickness of the 2DEG in a direction perpendicular 
to the plane of the 2DEG \cite{park,peters1,peters2,papic,mds,nuebler,kyril,yuli},
and Landau level mixing \cite{morf,ma,nuebler,quinn,toke,kyril,yuli}.
Morf and d'Ambrumenil assumed that $\delta^{int}$ is shared
for the two even denominator FQHSs therefore, based on Eq.(1), the two
energy gaps $\Delta^{meas}_{5/2}$ and $\Delta^{meas}_{7/2}$ both satisfy
\begin{equation}
    \Delta^{meas}=\delta^{int} E_C - \Gamma.  
\end{equation}
Within this model, the disorder broadening parameter $\Gamma$ is also shared by the $\nu=5/2$ and $\nu=7/2$ FQHSs
and independent measurements of $\Delta^{meas}_{5/2}$ and $\Delta^{meas}_{7/2}$ allow the extraction of the two parameters $\delta^{int}$ and $\Gamma$.  
This is achieved by plotting the two measured gaps against $E_C$ and 
by fitting a line to the two data points \cite{ma}; the slope of this line is $\delta^{int}$, 
whereas the intercept  with the vertical scale is $\Gamma$. 
Such an analysis is shown in Fig.4 for the three alloy samples exhibiting the $\nu=7/2$ FQHS and 
we summarized the parameters of the model in Table.I.  $\Delta^{meas}_{5/2}$
used in our analysis are from Ref.\cite{deng}.

\begin{figure}[t]
 \includegraphics[width=0.85\columnwidth]{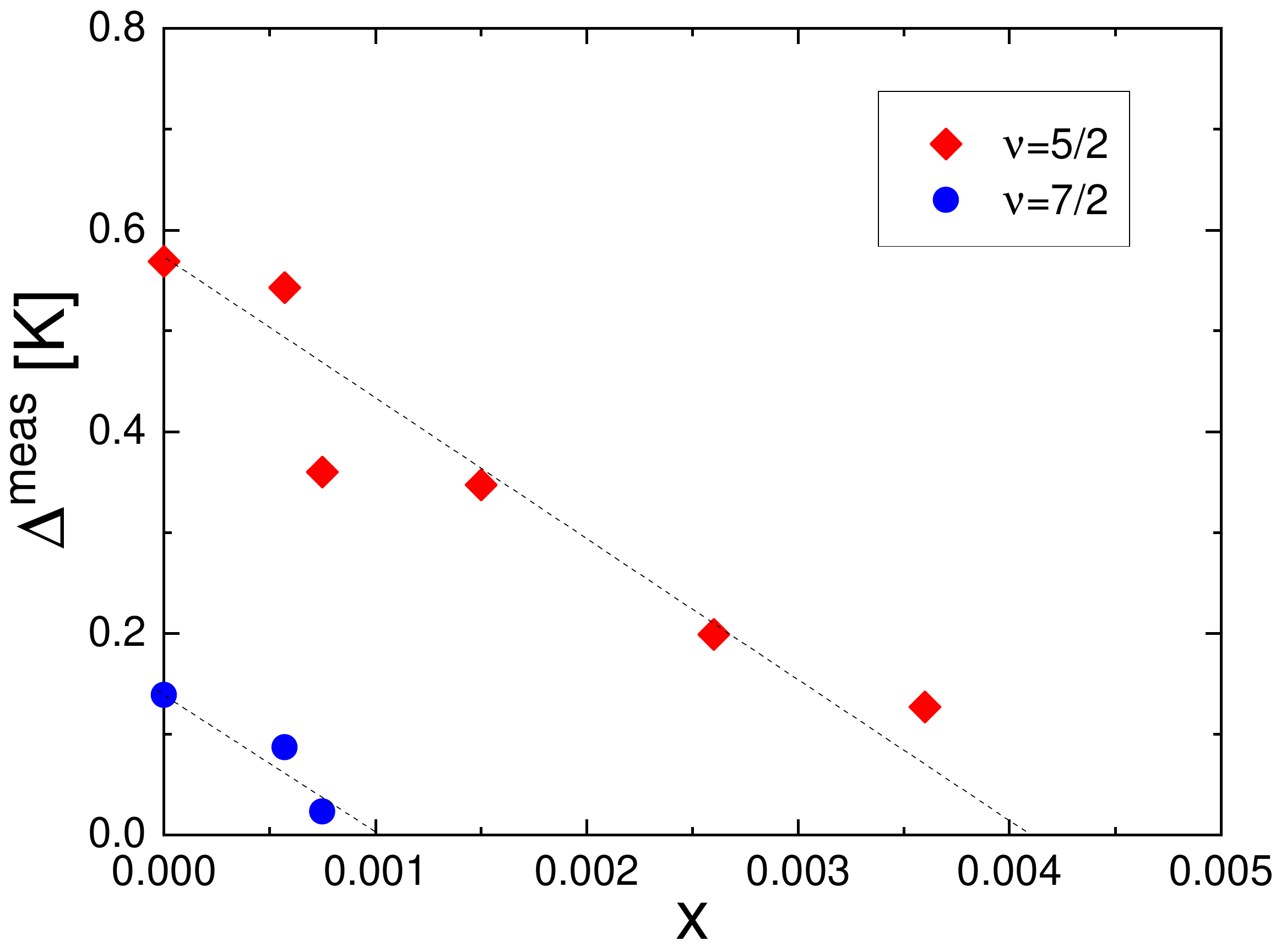}
 \caption{ Dependence of the measured energy gaps at $\nu=5/2$
 and $\nu=7/2$ on the alloy content $x$ of the channel. Energy gaps
 for the $\nu=5/2$ FQHS are from Ref.\cite{deng}. Dashed lines are guides to the eye.
  \label{Fig3}}
\end{figure}

\begin{table}[b]
\caption{A summary of the dimensionless intrinsic gap $\delta^{int}$ and 
disorder broadening parameter $\Gamma$ (in units of K) for the 
$\nu=5/2$ and $\nu=7/2$ FQHSs in a series of alloy samples. }
\begin{ruledtabular}
\begin{tabular}{l l l l l l l}
$x$     & 0 & 0.00057 & 0.00075 & 0.0015 & 0.0026 & 0.0036 \\
\hline
$\delta^{int}$ &  0.024 & 0.026 & 0.020 & - & - & - \\
$\Gamma$  & 2.21 & 2.40 & 1.82  & - & - & - \\
$\tilde{\Gamma}_{5/2}$  & 2.20 & 2.23 & 2.41  & 2.42 & 2.57 & 2.64 \\    
$\tilde{\Gamma}_{7/2}$  & 2.20 & 2.25 & 2.32  & - & - & - \\
\end{tabular}
\end{ruledtabular}
\end{table}

Analyses following the model of Morf and d'Ambrumenil
of the gaps in pristine samples from the literature
of electron density close to that of ours yielded
$\delta^{int} = 0.014$ \cite{ma} in a sample of $n=3.0 \times 10^{11}$~cm$^{-2}$ from Ref.\cite{eisen},  
$\delta^{int} = 0.019$ \cite{nodar} in a sample of $n=2.78 \times 10^{11}$~cm$^{-2}$ from Ref.\cite{nodar}, and
$\delta^{int} = 0.019$ \cite{nodar} in a sample of $n=3.0 \times 10^{11}$~cm$^{-2}$ from Ref.\cite{kumar}.
In our pristine sample $x=0$ we find $\delta^{int} = 0.024$, a value significantly
larger than those from the literature \cite{ma,nodar}. Nonetheless, this value is consistent with numerical
results as it is less than $\delta^{int} \approx 0.03$ \cite{feiguin,mds,kyril} and  $\delta^{int} \approx 0.036$ \cite{nuebler},
values obtained in the limit of no Landau level mixing and
zero layer thickness, but it is larger than
the values $\delta^{int} = 0.016$ \cite{ma}, 
$\delta^{int} = 0.018$ \cite{nuebler}, and $\delta^{int} = 0.016$ \cite{quinn}
from estimations that include both Landau level mixing and finite layer thickness effects.
However, such a consistency can only be considered crude at best because 
of systematic errors in our experiment discussed earlier and also because
the assumption of equal $\delta^{int}$ for the $\nu= 5/2$ and $7/2$
FQHSs is only approximate in measurements of samples of fixed density
due to the slightly different Landau level mixing parameters at these filling factors.

\begin{figure}[t]
 \includegraphics[width=1\columnwidth]{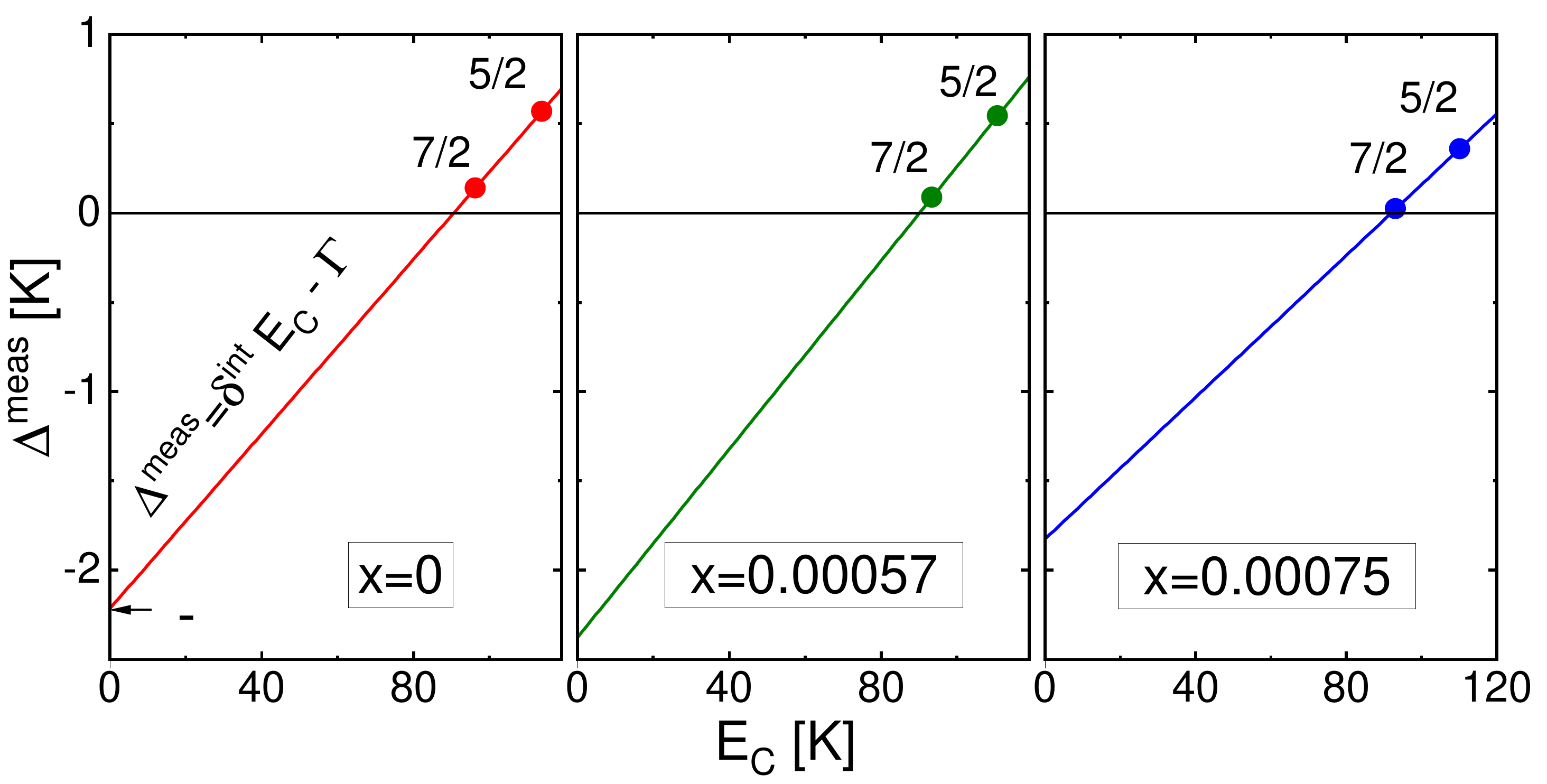}
 \caption{ Energy gaps $\Delta^{meas}_{5/2}$ and $\Delta^{meas}_{7/2}$ plotted 
 against the Coulomb energy $E_C$. According ideas put forth in Ref.\cite{ma},
 the slope of the line through the two data points is $\delta^{int}$, whereas the intercept of each line 
  with the vertical scale is the disorder broadening parameter $\Gamma$. 
 \label{Fig4}}
 \end{figure}
 
Since the series of alloy samples were engineered to have the same width of their
quantum well and have similar electron densities, their intrinsic gap is expected to be similar.
This is because within the phenomenological model embodied by Eqs.(1) 
alloy disorder factors only into the disorder broadening parameter, and not into the intrinsic gap.
Dimensionless intrinsic gaps of our alloy samples
listed in Table.I. are reasonably close, within a error of $\pm13\%$.
The sample with $x=0.00075$ has its $\delta^{int}$ the farthest from values in the other two samples.
We ascribe the scatter of $\delta^{int}$ to earlier discussed systematic errors.

As the amount of disorder increases, the measured energy gap is suppressed and
the disorder broadening parameter is expected
to increase. This trend can be observed  for samples
with $x=0$ and $x=0.00057$, but it breaks down as $x$ increases to $x=0.00075$.
This unphysical non-monotonic dependence of $\Gamma$ on $x$, shown in Fig.5, is likely
due to the same errors that led to variations in $\delta^{int}$. Because of 
multiple sources of errors, we are not able to disentangle the influence of different
sources or error on the disorder broadening parameter.
We conclude that, due to accumulating errors,
the model of Morf and d'Ambrumenil does not yield satisfactory disorder broadening parameters
in our alloy samples.

For an improved analysis we exploit the property of shared $\delta^{int}$ in
our series of samples. In the following we calculate a modified version of the
disorder broadening $\tilde{\Gamma}$ which is still based on Eq.(2), but in which
we fix $\delta^{int}=0.024$, its value
in our pristine sample. The parameter
$\tilde{\Gamma}=\delta^{int} E_C - \Delta^{meas}$
%Because of their different Coulomb energies,
%FQHSs at $\nu=5/2$ and $\nu=7/2$ have different intrinsic gaps
%$\Delta^{int}_{5/2}=2.77$ and  $\Delta^{int}_{5/2}=2.34$, respectively.
can be independently calculated at both $\nu=5/2$ and $\nu=7/2$. We label
these $\tilde{\Gamma}_{5/2}$ and $\tilde{\Gamma}_{7/2}$, list the obtained values in Table.I., and plot
these values against $x$ in Fig.5.
We found that both $\tilde{\Gamma}_{5/2}$ and $\tilde{\Gamma}_{7/2}$, when
plotted against $x$, exhibit significantly less scatter than $\Gamma$ and both
have an increasing trend with $x$, in agreement with our expectation.
Moreover, $\tilde{\Gamma}_{5/2}$ and $\tilde{\Gamma}_{7/2}$ are very close, 
with the largest difference of only $\pm2\%$. This shows that the modified analysis
is self-consistent and it is also consistent with the behavior shown in Fig.3
of similar $\delta \Delta^{meas}/\delta x$ slopes at $\nu=5/2$ and $\nu=7/2$.
We conclude that the modified analysis for obtaining the disorder
broadening parameter led to a significant improvement in the scatter of the data. 
%Such an analysis was enabled by engineering alloy samples with same quantum well width and similar density.

\begin{figure}[t]
 \includegraphics[width=1\columnwidth]{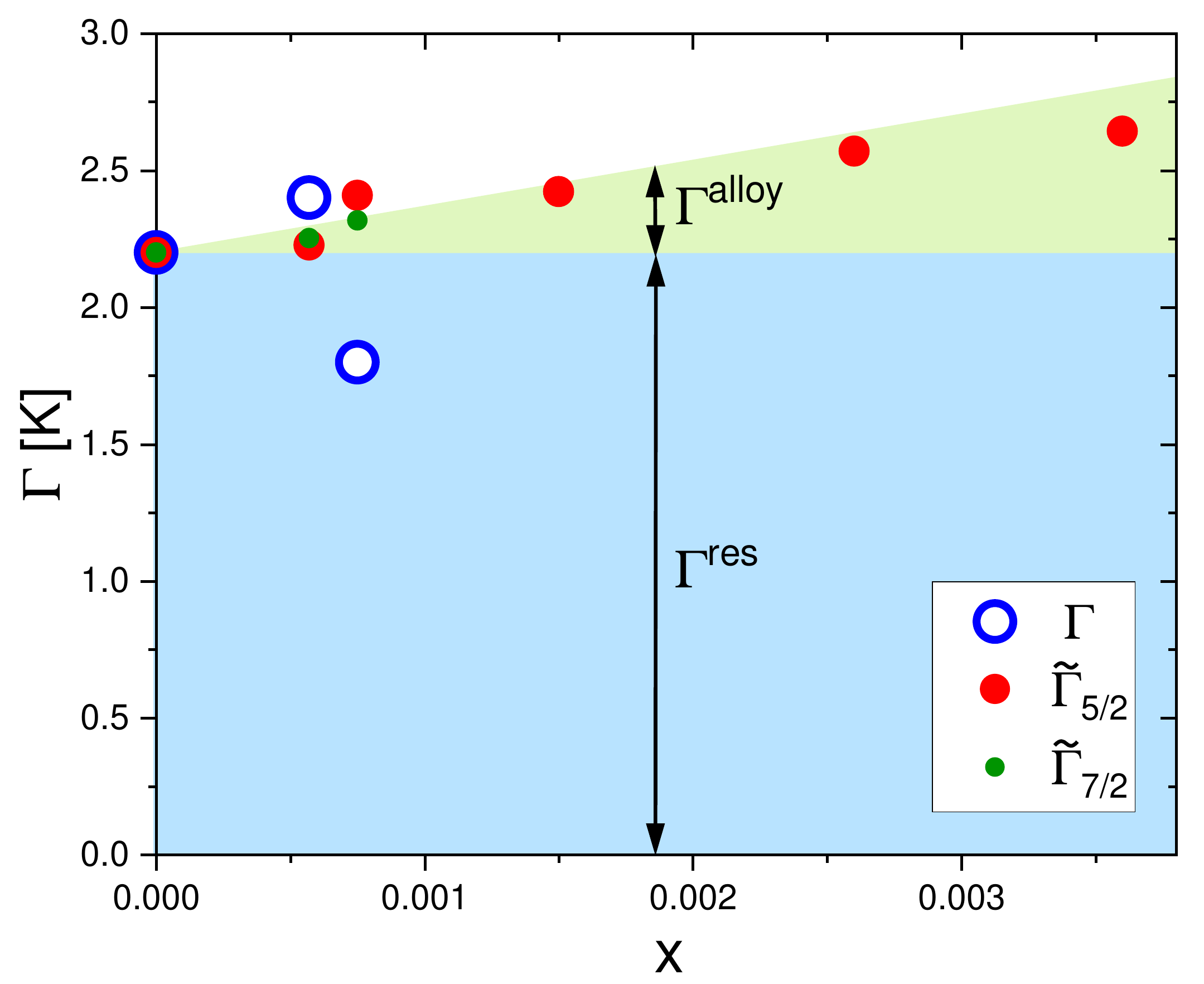}
 \caption{ Comparison of $x$ dependent disorder broadening parameters extracted using different models. 
 Open symbols represent $\Gamma$ extracted according to Ref.\cite{ma} while solid symbols are
 $\tilde{\Gamma}$ extracted with a fixed intrinsic gap at $\nu=5/2$ and $\nu=7/2$, as described in the text.
 \label{Fig5}}
\end{figure}

Fig.5 suggests that the disorder broadening parameter $\Gamma$ has two separate contributions:
one due to short range disorder scattering centers $\Gamma^{alloy}$ and another
due to the residual scattering $\Gamma^{res}$. The latter we associate with long range scattering
due to a smoothly varying scattering potential of ionized impurities present in all our samples;
$\Gamma^{res}$ is independent of the disorder level $x$. 
We suggest that  the disorder broadening admits the following
separation: $\Gamma = \Gamma^{res}+\Gamma^{alloy}(x)$.
Such a separation of the contributions of ionized impurities and alloy impurities
in our alloy samples is possible since they are grown under similar conditions,
thus share a similar long range scattering potential. 
We also found that for the range of alloy disorder studied, $\Gamma^{res} \gg \Gamma^{alloy}(x)$.
This finding quantifies earlier qualitative results, according to which
the long range potential due to ionized impurities is more detrimental to the
even denominator FQHSs than short range potentials \cite{pan11,deng}.
Our data yields $\Gamma^{alloy}(x)=0.17  \times 10^3 x$ in units of K.
Finally we note that the $\Gamma = \Gamma^{res}+\Gamma^{alloy}(x)$ relationship is reminiscent
of Matthiessen's rule for the scattering times: $1/\tau^{tot}=1/\tau^{res} + 1/\tau^{alloy}(x)$
found in disorder samples \cite{wanli0,gardner}. Here $\tau^{tot}$ is the scattering time
or mobility lifetime in alloy samples, $\tau^{res}$ is the scattering time due to the
ionized impurities, $\tau^{alloy}(x)$ is the alloy scattering time, which 
explicitly depends on the disorder level $x$.

Most recently, ideas were put forth according to which at these filling factors
disorder may induce microscopic puddles  \cite{ambrumenil,poly,d1,d2,d3,d4,d5}.
In Ref.\cite{ambrumenil}, in the presence of a long range scattering
due to remote ionized dopants the electron gas breaks up into puddles of
compressible and incompressible regions \cite{poly}
and the measured energy gap is determined by a so-called saddle-point energy gap
originating from thermally driven tunneling between the puddles \cite{ambrumenil}.
%Such a model is also applicable to scattering in high mobility
%pristine samples due to ionized impurities near or within the 2DEG.
%According to this model, 
%whose value is smaller than the intrinsic gap.
%Since the saddle-point energy gap depends only on the geometrical length scale of the potential landscape set by the ionized 
%impurities, the measured energy gap should be invariant for 
%samples of the same structure as well as the same doping scheme such as our set of alloy samples.
In this model \cite{ambrumenil} an inflection point in the Arrhenius plots of $\ln(R_{xx}$) versus $1/T$
needs to be established.
However, such a data analysis is fraught with difficulties because of a commonly occuring measurement
artifact. Indeed, this inflection point is expected to occur at low temperatures at which
electrons may not necessarily thermalize to the refrigerator
temperature, unless special thermalization techniques are used \cite{setup}. When electrons do not fully
thermalize, an inflection in the Arrhenius plots is observed which, however, is not necessarily related
to the puddling effect decribed in the model.
In addition, it is not clear whether this model is applicable for short range scattering potentials, such as the alloy
potential in our samples. Motivated by recent thermal Hall conductance measurements \cite{moty},
other theoretical work advocates for the formation 
of competing Pfaffian and anti-Pfaffian puddles on the microscopic scale \cite{d1,d2,d3,d4,d5}.
Within these models, the temperature dependence of the magnetoresistance and its relationship to the energery gap 
have not yet been worked out.

In conclusion, we have examined the energy gaps of the even denominator fractional quantum Hall
states in a series of alloy samples. Energy gaps in a series of alloy disorder samples
at both $\nu=5/2$ and $\nu=7/2$ are supressed similarly 
with alloy disorder.  In order to separate disorder and other effects, we used
an analysis of energy gaps proposed by Morf and d'Ambrumenil.
We found that the dimensionless intrinsic gaps are consistent with
numerical results, but are larger than those obtained from pristine samples of similar density
published in the literature.
Furthermore, the disorder parameter exhibited significant scatter. However, a modification of the model
enabled by a shared intrinsic gap yielded much improved results.
We found that the disorder broadening parameter may be split into contributions from
long-range scattering due to remote ionized impurities and that of short-range scattering due to alloy
disorder. The latter was found to increase linearly on the alloy content of our samples.

This work was supported by the DOE BES Experimental Condensed Matter Physics  
program under the Award No. DE-SC0006671.

\end{document}